# Near-Room-Temperature Field-Controllable Exchange Bias in 2D van der Waals Ferromagnet $Fe_3GaTe_2$

*Jifeng Shao[1,2,5], Xiaolong Yin[1,5], Chunhao Bao[1], Sirong Lu[1,2], Xiaoming Ma[1,2], Shu Guo[1,2], Le Wang[1,2], Xi Zhang[2], Zhiyue Li[1,3], Longxiang Li[2], Yue Zhao[1,2,4]✉, and Tingyong Chen[1,2]✉*

[1]*Shenzhen Institute for Quantum Science and Engineering, Southern University of Science and Technology, Shenzhen 518055, China.*
[2]*International Quantum Academy, Shenzhen 518048, China.*
[3]*Department of Physics, The University of of Hong Kong, Hong Kong, 999077, China.*
[4]*Department of Physics, Southern University of Science and Technology, Shenzhen 518055, China.*
[5]*These authors contributed equally: Jifeng Shao, Xiaolong Yin.*
✉ [zhaoy@sustech.edu.cn](mailto:zhaoy@sustech.edu.cn); [chenty@sustech.edu.cn](mailto:chenty@sustech.edu.cn).

**Abstract**

Exchange bias (EB) is a cornerstone of modern magnetic memory and sensing technologies. Its extension to the realm of two-dimensional (2D) van der Waals (vdW) magnets holds promise for revolutionary advancements in miniaturized and efficient atomic spintronic devices. However, the blocking temperature of EB in 2D vdW magnets is currently well below room temperature ~130 K. This study reports a robust EB phenomenon in $Fe_3GaTe_2$ thin-layer devices, which significantly increases the blocking temperature to a near-room-temperature record of 280 K. Both the bias direction and magnitude can be isothermally tuned by adjusting the field sweep range, in striking contrast to the conventional EB in ferromagnetic/antiferromagnetic (FM/AFM) bilayers. We propose an exchange spring model in which crystal defects with higher coercivity act as the pivotal pinning source for the observed EB phenomenon, deviating from the conventional FM/AFM interface mechanism. Cumulative growth of minor loops and multiple magnetization reversal paths are observed in field cycles below the saturation field, consistent with the hard FM defects behavior of our exchange spring model. These findings provide insights into the complex magnetic order in 2D ferromagnets and open new avenues for developing practical ultrathin vdW spintronic devices with EB-like properties at room temperature.

**Introduction**

Exchange bias (EB)[1, 2], initially identified at the interface of ferromagnetic/antiferromagnetic (FM/AFM) materials, is a manifestation of magnetic exchange coupling that induces a unidirectional shift of the FM hysteresis loop. EB is widely utilized in modern magnetic technologies, such as spin valve readback heads, MRAM memory circuits, and advanced disk media. Beyond the classical FM/AFM interfaces, EB phenomena, characterized by similar unidirectional shifts driven by magnetic exchange anisotropy, have been observed in other configurations containing uncompensated spins[3, 4] at the interface of the free and pinning layer, including systems with ferrimagnets[5], harder ferromagnets (compared to the free layer)[6, 7, 8], or spin glasses[9, 10] serving as the pinning layer. Elucidating the underlying mechanisms of EB and its analogous phenomena across diverse magnetic systems not only illuminates fundamental magnetic interactions but also unlocks new possibilities for spintronic advancements[11].

The remarkable perpendicular magnetic anisotropy (PMA) observed in 2D van der Waals (vdW) magnets, even down to the monolayer scale[12, 13, 14, 15, 16], underscores their potential in the next-generation ultrathin spintronic devices. The strong ferromagnetic intralayer and weak interlayer exchange coupling provide an ideal arena for manipulating magnetic ground states through various methods, such as compressing the interlayer space[17, 18], intercalating magnetic atoms into vdW gaps[10, 19], proton gating[20, 21], and modulating stacking order[22, 23, 24, 25, 26]. The tunability, coupled with the rich magnetic phases inherent to 2D vdW materials and their heterostructures, provides a fertile ground for the exploration of the EB phenomenon. Traditional designs with AFM/FM interface in artificial 2D vdW heterostructures often face challenges from weak interfacial exchange coupling and limited magnetic transition temperatures, resulting in low bias field ($H_{ex}$) and blocking temperature ($T_B$)[27]. On the other hand, the exploration in singular vdW 2D magnets unveils a diverse origin of EB phenomena because of their complex magnetic order. Examples include the coexistence of AFM and spin glass order in $Fe_xNbS_2$[10], atomic anti-site disorder in ferrimagnetic monolayer $MnSb_2Te_4$[15], and AFM/FM interfacial contributions in $MnBi_4Te_7$[28], $MnBi_6Te_{10}$[28], and $Fe_3GeTe_2$ (FGT)[20].

Despite the discovery of vdW ferromagnets with $T_C$ above room temperature and strong PMA[29, 30, 31], the realization of EB in these systems has been constrained by blocking temperature significantly below room temperature, approximately 130 K[18, 27], limiting practical applications. Herein, we report a robust EB phenomenon with a near-room temperature $T_B$ (~ 280 K) and a sizable $H_{ex}$ (0.25 T at 1.8 K) in thin-layer $Fe_3GaTe_2$ (FGaT) devices. The direction and magnitude of $H_{ex}$ can be tuned isothermally by varying the field sweep ranges without requiring a field cooling procedure. We propose a "hard/soft FM" exchange spring model in which crystal defects with higher coercivity serve as the hard pinning source. Furthermore, the FGaT devices exhibit a cumulative growth of minor loops and multiple magnetization reversal paths during magnetic field cycles within the saturation field. This agrees with the presence of hard FM defects in the proposed ES model from another perspective. The EB is durable and shows no signs of attenuation or annihilation during 86-cycle training measurements. Our findings provide new insight into the EB mechanism in singular 2D vdW magnets. The durable and sizable EB phenomenon, which persists up to near room temperature, marks a pivotal advancement in expanding the potential of 2D vdW magnets for room-temperature spintronic applications.

**Results**

**Structural characterization and magnetic properties of $Fe_3GaTe_2$**

FGaT single crystals were grown by the self-flux method and characterized by single-crystal X-ray diffraction (Fig. S1) and scanning transmission electron microscopy (STEM) (Fig. 1b). The crystals exhibit a layered hexagonal crystal structure in which $Fe_3Ga$ slabs are covalently bonded and sandwiched by two Te layers (Fig. 1a). The air sensitivity of FGaT nanoflakes is evidenced by the surface oxidation layers observed in the atomic-resolution high-angle annular dark-field (HAADF) images of the $(1\bar{2}0)$ crystal plane in Fig. 1b. Both our bulk single crystal and thin-layer device #1 (146 nm) exhibit a Curie temperature ($T_C$) of ~ 340 K, consistent with previous reports[16, 30]. The $R_{xy}$-$T$ curves of device #1 agree with the $M$-$T$ curves of the bulk single crystal (Fig. 1d, Fig. S2), as $R_{xy}$ is proportional to the magnetic moment $M$. However, the remanent $R_{xy}$ of device #1 sharply decreases at around 300 K for cooling fields above 0.08 T, while the remanent $M$ of the bulk single crystal decreases smoothly (Fig. S3b-3c). This suggests that the bulk and thin-layer samples possess different domain structures, and the thin-layer device #1 undergoes a pronounced depinning effect associated with a hard-to-soft domain structure transition at around 300 K. The recent report shows that the soft domain width is one order of magnitude smaller than the width of Hall electrodes (3 μm)[32]. Similar phenomena have been reported in FGT thin-layer devices[33, 34]. Device #1 displays typical pinning ferromagnets behavior with multidomain structures in the anomalous Hall effect (AHE) hysteresis loops (Fig. 1e). The initial $R_{xy}^A$ ($\mu_0H = 0$) is negligible and changes little at first, but then undergoes a steep jump to the saturation value as the magnetic field increases. The saturation field of the virgin curve is much smaller than that of the overall hysteresis loop, indicating an additional pinning effect. A rarely seen zero field cooled (ZFC) EB phenomenon ($\mu_0H_{ex} = \mu_0(H_c^+ + H_c^-)/2 = 0.095$ T) is detected in the AHE hysteresis loops. The bias direction appears to be determined by the virgin magnetization direction, as observed in $Mn_2PtGa$[35] and $InMnNi_{13}$[36]. However, the bias direction of the ZFC EB is random in subsequent remeasurements (Fig. S4).

**Near-room-temperature EB phenomenon in Fe₃GaTe₂ thin-layer device**

The field-cooled (FC) EB phenomenon was measured by with applied fields of +3 T and -3 T, respectively (Fig. 2). Device #1 consistently exhibits a negative EB below 280 K ($\mu_0H_{ex}$ (270 K) ~ 0.012 T). The near-room temperature $T_B$ exceeds twice the current highest record (130 K) in vdW systems[18, 27]. Above 200 K, the sharp spin reversal is incomplete, accompanied by roughly linear tails on the hysteresis loops (Fig. 2 and Fig. S5). With increasing temperature, the abrupt jumps gradually diminish and shift towards the initial saturation field direction, while the linear part becomes more prominent. The remanent $R_{xy}$ ($\mu_0H = 0$) drops to zero at temperatures above 300 K (Fig. S5), consistent with the depinning temperature on the remanent $R_{xy}$-$T$ curves (Fig. 1d and Fig. S3b). The thermal evolution of the hysteresis loop shape is attributed to the reduced PMA due to thermal fluctuation and the emergence of labyrinthine stripe domains at intermediate temperatures[16, 30, 32, 37]. The stripe domains, along with topological magnetic bubbles, were captured in the Lorentz transmission electron microscopy (LTEM) image of the FGaT nanoflake at 220 K (Fig. S6).

**Field-controllable EB and the proposed exchange spring model**

Given the intricate nature of magnetic order and its impact on mesoscopic transport, the underlying mechanism of this near-room-temperature EB phenomenon could be complicated. To further investigate the type of pinning sources, we analyze the EB response to different field sweep ranges. It was discovered that the direction and magnitude of the $H_{ex}$ can be isothermally controlled by adjusting the field sweep ranges, in addition to using the FC process. The $H_{ex}$ value increases monotonically as the absolute value of the negative field set (NFS) increases, while the positive field set (PFS) is fixed at 1 T (Fig. 3a). The $\mu_0H_{ex}$ value is significantly enhanced from 0.11 T to 0.25 T when the field sweep range of measurement changes from (-1 T, 1 T) to (-14 T, 1 T) (Fig. 3b). When the direction of initial saturation is switched, the bias direction reverses. For example, the bias directions of the negative protocol (NFS>PFS), such as (-2/-14 T, 1 T), and positive protocol (NFS<PFS), such as (-1T, 2/14 T), are opposite (Fig.3b, Fig. S7). Moreover, the $\mu_0H_{ex}$ disappears but the $\mu_0H_C$ (0.93 T) almost doubles the initial value (0.58 T) when measured by a symmetric protocol (NFS=PFS) with large field ranges (14 T).

In FM/AFM bilayers, the bias direction of $H_{ex}$ is typically controlled by the cooling field applied from temperature above the Néel temperature ($T_N$) of AFM, which is otherwise difficult to change at low temperatures because the AFM is generally insensitive to external magnetic fields. The $T_B$ of EB is lower than the $T_c$ of FM and the $T_N$ of AFM. Therefore, it is unlikely that the pinning source originates from a possible AFM surface oxidation layer, as the AFM exchange coupling of Fe-O with $T_N$ above 270 K shall be very strong at 1.8 K. Furthermore, a pronounced EB phenomenon ($T_B$ above 250 K) was observed on the reflective magnetic circular dichroism (RMCD) hysteresis loops of freshly exfoliated FGaT nanoflakes that were protected by pure argon throughout the entire process (Fig. 3d, Fig. S8). The EB phenomenon observed in RMCD can also be isothermally controlled by adjusting field sweep ranges. No second harmonic generation signal was detected, indicating the absence of an AFM signal in the fresh FGaT nanoflakes. When the AFM interlayer coupling is weak and can be fully polarized by the sweeping fields, the unidirectional pinning effect fails, and EB vanishes[28, 38]. In this senario, the saturation field of the possible AFM should lie between 1 T and 2 T, as indicated by the opposite bias directions observed when using the positive protocol (-1 T, 2 T) and negative protocol (-2 T, 1 T) (Fig. S7). However, it becomes challenging to account for the observed increase in $H_{ex}$ when using more negative protocols such as (-3/-4/-5 T, 1 T) in Fig. 3a, as well as the $H_{ex}$ values obtained using symmetric protocol of 5 T in Fig. S7.

On the other hand, the sign of EB in AFM/FM bilayers is primarily determined by the type of interfacial coupling and the magnitude of the cooling field[39]. FM-type interfacial coupling always results in a negative EB, while AFM-type interfacial coupling usually shows a negative EB when the cooling field is low. As the cooling field increases, the negative EB gradually decreases and eventually becomes positive when the Zeeman energy in the AFM layer exceeds the AFM coupling energy between the AFM and FM layers[39, 40, 41, 42]. Notably, the sign of $H_{ex}$ at 220 K reverses three times suddenly without any transition trace as the cooling field increases from 0 to 0.25 T. Moreover, it remains negative for cooling fields larger than 0.25 T (Fig. 3c, Fig. S9), indicating that the pinning source is not the possible AFM phase.

In the "soft/hard FM" exchange spring magnets, the bias direction of the soft FM hysteresis loop reverses when the hard FM is switched by a large field. Additionally, the field sweep range can manipulate the bias value, which depends on the absolute value of the hard FM magnetization [43]. Therefore, the pinning source is likely the hard FM, as illustrated in Fig. 3e-f. The interfacial coupling between the hard and soft FGaT should be FM-type, as indicated by the consistently negative EB obtained under large cooling field and asymmetric field sweep protocols. The high magnetic field in the asymmetric protocols (Fig. 3e and Fig. 3g) aligns the soft and hard spins, while the low opposite field fails to switch the hard spins. Thus, the pinning effect is unidirectional, and the hysteresis loop shifts to the low-field side. The absolute value of the $H_{ex}$ increases when the starting magnetic field polarizes more hard spins and the reversed field switches fewer hard spins. If the field range of symmetric protocol is large enough to fully polarize all spins (Fig. 3f), the hard pinning effect in the two field sweep branches becomes symmetric, resulting in a symmetric hysteresis loop with an enlarged $H_c$. However, if the field range of symmetric protocol is small to switch all the hard spins, a biased hysteresis loop may also be observed, and the bias direction is determined by the orientations of hard spins. Therefore, the sign of $H_{ex}$, as measured under the ZFC process and low cooling field, is initially random but remains constant during the subsequent repeated measurements (Fig. 4 and Fig. S9-10). The complete hysteresis loops of the soft/hard FM bilayers generally display well-separated magnetization reversals of the soft and hard layers[8, 43]. However, the spin switching of hard FM is unobservable in the symmetric hysteresis loops of AHE and RMCD. Therefore, the hard FM is probably related to the crystal defects that induce a spatial distribution of different coercive fields, and their sizes are much smaller than the optical resolution of RMCD instruments. Similar results have been reported in monolayer $MnSb_2Te_4$[15].

**The remarkable durability of EB**

Conventional three-dimensional FM/AFM and soft/hard FM heterostructure devices often exhibit an aging effect, where their $H_{ex}$ experiences a significant decrease from the first to the second cycle, followed by a gradual decay in the subsequent cycles[43, 44]. In contrast, the EB phenomenon in FGaT devices shows little sign of attenuation or annihilation in the consecutive 86-cycle repeated measurements (Fig. 4). This remarkable durability is valuable for practical applications. It is noteworthy that $H_c$ and $H_{ex}$ fluctuate around their initial values and do not show a specific correlation with each other in the 86-cycle training measurements. The durability and fluctuation are attributed to the spatial distribution of harder defects with different coercivities, which can induce unstable nucleation[16, 37] and multiple metastable microscopic pathways for magnetization reversal[19, 45].

**The cumulative growth of minor loops and field-controllable EB at intermediate temperature**

We also observed a cumulative growth of minor hysteresis loops in the FGaT device #2 when measured using the symmetric protocol of 0.2 T at 200 K (see the black circle in Fig. 5a). This phenomenon is typically observed in thin Co/Pt films with a spatial distribution of hard defects that cause an asymmetric response to the reversal of the sweeping field[46, 47]. This observation agrees with our proposed exchange spring model with the presence of harder defects. The minor loops rapidly reached saturation after only one cycle, as evidenced by the equal saturation value of hysteresis loops measured by larger field sweep ranges, such as 0.25 T and 3 T. Notably, the saturation remains constant when the symmetric field sweep range is reduced to 0.19 T. After five repeated cycles measured by the SP of 0.25 T (Fig. S12), we conducted an additional 20 cycles using the SP of 0.2 T. Unlike the consistent shapes of the repeated cycles of 0.25 T, a variety of hysteresis loops with different shapes but constant saturation values randomly emerged (Fig. S13). Five representative loops are selected and shown in Fig. 5b. The consistent saturation values underscore that the magnetization state of the pinning source is a key factor influencing the shape of hysteresis loops. The uniform magnetization of hard FM is crucial for the EB in exchange spring magnets[43]. At 200 K, thermal fluctuations reduce the PMA of both soft and hard FGaT. The orientations of hard spins become random as a magnetic field of 0.2 T cannot align them. Additionally, the emergence of stripe domains also contribute to the dynamic interfacial spin configurations during the field sweeping[48], resulting in the random loops (Fig. 5b). Initiating the field sweep from a specific uniform magnetization state enables the selective control over EB with correlated bias direction (Fig. 5c).

Based on our results and analysis, caution should be exercised when attributing the EB phenomenon in singular ferromagnets with strong PMA (such as FGaT, FGT, VI$_3$[49], and Co$_3$Sn$_2$S$_2$[50]) to a potential coexisting AFM phase. To identify the pinning source, a systematic investigation that includes the response to different field sweep ranges is essential. We observed a similar EB phenomenon that can be controlled isothermally by different field sweep ranges in our FGT thin-layer devices (Fig. S15). Therefore, conflicting experimental reports on the exchange bias phenomenon in oxidized FGT devices, such as its absence in recent AHE[51] and RMCD measurements[18], may arise from the different quality of FGT crystal used by different groups rather than the surface oxidation layers. It is intriguing to note that the techniques reported to induce EB phenomenon in thin-layer FGT, such as ion irradiation[52], proton intercalation[21], interlayer spacing compression[18], and excessive Fe-intercalation into the vdW gap[19], all involve the introduction of inhomogeneous disorders. This raises the question of how these EBs respond to different field sweep ranges, warranting further investigation. Moreover, our findings suggest that introducing harder coercivity in vdW ferromagnets with strong PMA and a $T_C$ above room temperature may present an opportunity for achieving the long-sought ultrathin vdW spintronics with room-temperature EB-like properties.

**Discussion**

In summary, we have observed a field-controllable EB in the FGaT thin-layer devices, and the EB temperature in a vdW system is substantially elevated from the current 130 K to near-room temperature at 280 K. The magnitude and direction of the EB can be controlled by adjusting the field-sweep range and direction without going above the blocking temperature such as those FC in conventional EB structures. The EB is durable and shows little signs of attenuation or annihilation after 86 cycles of repeated measurements. We propose an exchange spring model, in which randomly distributed crystal defects with higher coercivity act as the pinning source, to explain the observed behaviors that deviate from the conventional EB of FM/AFM bilayers. Furthermore, the presence of hard FM defects in our exchange spring model is supported by the cumulative growth of minor loops and the multiple magnetization reversal paths observed during field cycles within the saturation field. These findings provide valuable insights into the nature of EB phenomena in singular ferromagnets with strong PMA and open new avenues for the development of practical ultrathin vdW spintronics with room-temperature EB-like properties.

**Methods**
**Crystal growth**

Single crystals of Fe$_3$GaTe$_2$ were grown by the self-flux method in muffle furnace. High purity Fe powder (99.99%), Ga lump (99.98%) and Te powder (99.99%) were sealed in a vacuum quartz tube with a molar ratio of 1:1:2 (Fe: Ga: Te). The mixture was heated up to 1000 °C and held for 24 hours. It was then rapidly cooled down to 780 °C within 1 hour and kept for another 100 hours. The platelet-like single crystals were then separated from the excess flux by high temperature centrifugation.

**Single-crystal x-ray diffraction**

The crystal quality and structure of the FGaT single crystals were first characterized by the single-crystal x-ray diffraction (SCXRD) patterns at 300 K with a Kappa Apex2 charge-coupled device diffractometer (Bruker D8 VENTURE) using graphite-monochromatized Mo-Kα radiation (λ = 0.71073 Å). Raw data were corrected for polarization, background, Lorentz factor, and multi-scan absorption, using APEX3 software and the SADABS-2016/2 program package. The crystal structure was refined using the SheLXL least-squares refinement package within the Olex2 program. The SCXRD patterns of FGaT in three different reciprocal lattice planes ($0kl$, $h0l$ and $hk0$) and detailed crystal structure information are shown in Fig. S1 and Tables S1-S2.

**STEM and LTEM characterization**

The FGaT single crystal was mechanically exfoliated into nanoflakes and then transferred onto $SiO_2$/Si substrate. The FGaT nanoflakes were capped with Pt layer to avoid further exposure to air and ion damage. Then, the FGaT nanoflakes were cut by a focused ion beam (FIB) system along two crystal planes of (100) and ($1\bar{2}0$), respectively. The thin wedge nanoflakes used for LTEM measurements were cut from the single crystal along the (001) plane and thinned by FIB. Atomic resolution HADDF-STEM images were obtained on a double Cs-corrected transmission electron microscope (Titan Themis G2) with an accelerating voltage of 300 kV. The domain structures were observed by using an ambient Cs-corrected transmission electron microscope (Titan ETEM G2) equipped with low temperature holder (liquid nitrogen, ~77 K) in the low magnification Lorentz mode, which can apply a normal field ranging from -2000 Oe to 2000 Oe.

**Magnetic and transport measurements**

The single crystal $Fe_3GaTe_2$ was first identified by the magnetic properties using MPMS3 and then mechanically exfoliated into thin-layer nanoflakes with silicon-free blue tape and then transferred onto $SiO_2$/Si substrate. The standard Hall bar electrodes were patterned by a laser direct writing machine (DWL 66+) and then coated with Ti/Au(5nm/100nm) using an electron beam evaporation coating system (JEB-2). The thickness of the device was measured by atomic force microscopy (AFM, Oxford). All transport measurements were performed in a physical property measurement system (PPMS DynaCool) with a base temperature of 1.8 K and a magnetic field of up to 14 T.

**Magneto-optical measurements of the pure FGaT nanoflakes**

The surface oxidation layers of the FGaT single crystal were first removed with blue tapes. Pure FGaT nanoflakes were mechanically exfoliated on $SiO_2$/Si substrate, and then transferred in a specially designed sealed box with an optical window in the center of the lid. All these processes were performed in an argon-filled glove box with very low oxygen ($O_2$ < 0.1 ppm) and moisture ($H_2O$ < 0.1 ppm). The sealed box filled with pure argon was then quickly transferred to the high vacuum chamber of the cryostat within half a minute. After cooling down to the set temperature, the in-situ MCD and SHG measurements were performed immediately. A 632.8nm He-Ne laser is used as the light source and the polarization state of the incident light beam changes between left- and right-handed circular polarization with a frequency of 50 KHz in the RMCD measurements. The SHG spectrum was excited by an 800 nm femtosecond laser generated from a Ti-Sapphire oscillator (Chameleon Ultra II) with an 80 MHz repetition and 150 fs pulse width. After passing through a long working distance objective, the light spot on the sample is ~2 μm.

**References**

1.  Meiklejohn W. H., Bean C. P. New Magnetic Anisotropy. *Phys. Rev.* **102**, 1413-1414 (1956).

2.  Nogués J., Schuller I. K. Exchange bias. *J. Magn. Magn. Mater.* **192**, 203-232 (1999).

3.  Takano K., Kodama R. H., Berkowitz A. E., Cao W., Thomas G. Interfacial Uncompensated Antiferromagnetic Spins: Role in Unidirectional Anisotropy in PolycrystallineNi81Fe19/CoOBilayers. *Phys. Rev. Lett.* **79**, 1130-1133 (1997).

4.  Ohldag H., *et al.* Correlation between exchange bias and pinned interfacial spins. *Phys. Rev. Lett.* **91**, 017203 (2003).

5.  Nayak A. K., *et al.* Design of compensated ferrimagnetic Heusler alloys for giant tunable exchange bias. *Nat. Mater.* **14**, 679-684 (2015).

6.  Berger A., Hovorka O., Friedman G., Fullerton E. E. Nonlinear and hysteretic exchange bias in antiferromagnetically coupled ferromagnetic bilayers. *Phys. Rev. B* **78**, 224407 (2008).

7.  Fullerton E. E., Jiang J. S., Grimsditch M., Sowers C. H., Bader S. D. Exchange-spring behavior in epitaxial hard/soft magnetic bilayers. *Phys. Rev. B* **58**, 12193-12200 (1998).

8.  Fullerton E. E., Jiang J. S., Bader S. D. Hard/soft magnetic heterostructures: model exchange-spring magnets. *J. Magn. Magn. Mater.* **200**, 392-404 (1999).

9.  Ali M., *et al.* Exchange bias using a spin glass. *Nat. Mater.* **6**, 70-75 (2007).

10. Maniv E., *et al.* Exchange bias due to coupling between coexisting antiferromagnetic and spin-glass orders. *Nat. Phys.* **17**, 525-530 (2021).

11. Phan M.-H., *et al.* Exchange bias and interface-related effects in two-dimensional van der Waals magnetic heterostructures: Open questions and perspectives. *J. Alloys Compd.* **937**, (2023).

12. Huang B., *et al.* Layer-dependent ferromagnetism in a van der Waals crystal down to the monolayer limit. *Nature* **546**, 270-273 (2017).

13. Gong C., *et al.* Discovery of intrinsic ferromagnetism in two-dimensional van der Waals crystals. *Nature* **546**, 265-269 (2017).

14. Lin Z., *et al.* Magnetism and Its Structural Coupling Effects in 2D Ising Ferromagnetic Insulator $VI_3$. *Nano Lett.* **21**, 9180-9186 (2021).

15. Zang Z., *et al.* Exchange Bias Effects in Ferromagnetic $MnSb_2Te_4$ down to a Monolayer. *ACS Appl. Electron. Mater.* **4**, 3256-3262 (2022).

16. Chen Z., Yang Y., Ying T., Guo J.-g. High-$T_c$ Ferromagnetic Semiconductor in Thinned 3D Ising

Ferromagnetic Metal Fe$_3$GaTe$_2$. *Nano Lett.* **24**, 993-1000 (2024).

17. Shao J., *et al.* Pressure-Tuned Intralayer Exchange in Superlattice-Like MnBi$_2$Te$_4$/(Bi$_2$Te$_3$)$_n$ Topological Insulators. *Nano Lett.* **21**, 5874-5880 (2021).

18. Liu C., *et al.* Emergent, Non-Aging, Extendable, and Rechargeable Exchange Bias in 2D Fe$_3$GeTe$_2$ Homostructures Induced by Moderate Pressuring. *Adv. Mater.* **35**, e2203411 (2022).

19. Wu Y., *et al.* Fe-Intercalation Dominated Ferromagnetism of van der Waals Fe$_3$GeTe$_2$. *Adv. Mater.* **35**, 2302568 (2023).

20. Zheng G. L., *et al.* Gate-Tuned Interlayer Coupling in van der Waals Ferromagnet Fe$_3$GeTe$_2$ Nanoflakes. *Phys. Rev. Lett.* **125**, 047202 (2020).

21. Wang C. S., *et al.* Sign-tunable exchange bias effect in proton-intercalated Fe$_3$GaTe$_2$ nanoflakes. *Phys. Rev. B* **107**, L140409 (2023).

22. Liang S. C., *et al.* New coercivities and Curie temperatures emerged in van der Waals homostructures of Fe$_3$GeTe$_2$. *Phys. Rev. Mater.* **7**, L061001 (2023).

23. Xu Y., *et al.* Coexisting ferromagnetic-antiferromagnetic state in twisted bilayer CrI$_3$. *Nat. Nanotechnol.* **17**, 143-147 (2022).

24. Song T., *et al.* Switching 2D magnetic states via pressure tuning of layer stacking. *Nat. Mater.* **18**, 1298-1302 (2019).

25. Li T., *et al.* Pressure-controlled interlayer magnetism in atomically thin CrI$_3$. *Nat. Mater.* **18**, 1303-1308 (2019).

26. Chen W., *et al.* Direct observation of van der Waals stacking–dependent interlayer magnetism. *Science* **366**, 983-987 (2019).

27. Huang X., *et al.* Manipulating exchange bias in 2D magnetic heterojunction for high-performance robust memory applications. *Nat. Commun.* **14**, 2190 (2023).

28. Xu X., *et al.* Ferromagnetic-antiferromagnetic coexisting ground state and exchange bias effects in MnBi$_4$Te$_7$ and MnBi$_6$Te$_{10}$. *Nat. Commun.* **13**, 7646 (2022).

29. Wang H., *et al.* Interfacial engineering of ferromagnetism in wafer-scale van der Waals Fe$_4$GeTe$_2$ far above room temperature. *Nat. Commun.* **14**, 2483 (2023).

30. Zhang G., *et al.* Above-room-temperature strong intrinsic ferromagnetism in 2D van der Waals Fe$_3$GaTe$_2$ with large perpendicular magnetic anisotropy. *Nat. Commun.* **13**, 5067 (2022).

31. Ma X., *et al.* Ferromagnetism above Room Temperature in Two Intrinsic van der Waals Magnets

with Large Coercivity. *Nano Lett.* **23**, 11226-11232 (2023).

32. Zhang G., *et al.* Field-free room-temperature modulation of magnetic bubble and stripe domains in 2D van der Waals ferromagnetic $Fe_3GaTe_2$. *Appl. Phys. Lett.* **123**, 101901 (2023).

33. Tan C., *et al.* Hard magnetic properties in nanoflake van der Waals $Fe_3GeTe_2$. *Nat. Commun.* **9**, 1554 (2018).

34. Fei Z., *et al.* Two-dimensional itinerant ferromagnetism in atomically thin $Fe_3GeTe_2$. *Nat. Mater.* **17**, 778-782 (2018).

35. Nayak A. K., *et al.* Large zero-field cooled exchange-bias in bulk $Mn_2PtGa$. *Phys. Rev. Lett.* **110**, 127204 (2013).

36. Wang B. M., *et al.* Large exchange bias after zero-field cooling from an unmagnetized state. *Phys. Rev. Lett.* **106**, 077203 (2011).

37. Jagla E. A. Hysteresis loops of magnetic thin films with perpendicular anisotropy. *Phys. Rev. B* **72**, 094406 (2005).

38. Zhu R., *et al.* Exchange Bias in van der Waals $CrCl_3/Fe_3GeTe_2$ Heterostructures. *Nano Lett.* **20**, 5030-5035 (2020).

39. Nogués J., Leighton C., Schuller I. K. Correlation between antiferromagnetic interface coupling and positive exchange bias. *Phys. Rev. B* **61**, 1315-1317 (2000).

40. Wang F., *et al.* Observation of Interfacial Antiferromagnetic Coupling between Magnetic Topological Insulator and Antiferromagnetic Insulator. *Nano Lett.* **19**, 2945-2952 (2019).

41. Nogués J., Lederman D., Moran T. J., Schuller I. K. Positive Exchange Bias in $FeF_2$-Fe Bilayers. *Phys. Rev. Lett.* **76**, 4624-4627 (1996).

42. Ying Z., *et al.* Large Exchange Bias Effect and Coverage-Dependent Interfacial Coupling in $CrI_3/MnBi_2Te_4$ van der Waals Heterostructures. *Nano Lett.* **23**, 765-771 (2023).

43. Binek C., Polisetty S., He X., Berger A. Exchange bias training effect in coupled all ferromagnetic bilayer structures. *Phys. Rev. Lett.* **96**, 067201 (2006).

44. Hoffmann A. Symmetry Driven Irreversibilities at Ferromagnetic-Antiferromagnetic Interfaces. *Phys. Rev. Lett.* **93**, 097203 (2004).

45. Pierce M. S., *et al.* Disorder-induced microscopic magnetic memory. *Phys. Rev. Lett.* **94**, 017202 (2005).

46. Berger A., Mangin S., McCord J., Hellwig O., Fullerton E. E. Cumulative minor loop growth in

Co/Pt and Co/Pd multilayers. *Phys. Rev. B* **82**, 104423 (2010).

47. Windsor Y. W., Gerber A., Karpovski M. Dynamics of successive minor hysteresis loops. *Phys. Rev. B* **85**, 064409 (2012).

48. Pierce M. S., *et al.* Quasistatic X-Ray Speckle Metrology of Microscopic Magnetic Return-Point Memory. *Phys. Rev. Lett.* **90**, 175502 (2003).

49. Zhang X., *et al.* Strain Tunability of Perpendicular Magnetic Anisotropy in van der Waals Ferromagnets $VI_3$. *Nano Lett.* **22**, 9891-9899 (2022).

50. Noah A., *et al.* Tunable exchange bias in the magnetic Weyl semimetal $Co_3Sn_2S_2$. *Phys. Rev. B* **105**, 144423 (2022).

51. Kim D., *et al.* Antiferromagnetic coupling of van der Waals ferromagnetic $Fe_3GeTe_2$. *Nanotechnology* **30**, 245701 (2019).

52. Wu Q., *et al.* Giant and Nonvolatile Control of Exchange Bias in $Fe_3GeTe_2$/Irradiated $Fe_3GeTe_2$/MgO Heterostructure Through Ultralow Voltage. *Adv. Funct. Mater.* **33**, 2214007 (2023).

**Acknowledgements**
The authors acknowledge support from the National Key R&D Program of China (Grants No. 2022YFA1403700), the National Natural Science Foundation of China (Grants Nos. 11974158, 11804402, 12204221, 12204223, 22205091, 52103282), the Key-Area Research and Development Program of Guangdong Province (Grants Nos. 2020B0303050001, 2021B0101300001), the Guangdong Basic and Applied Basic Research Foundation (Grants No. 2024A1515030118, 2022A515012283), the Innovative Team of General Higher Educational Institutes in Guangdong Province (Grants No. 2020KCXTD001). S.L. acknowledges the support of Pico Center at Southern University of Science and Technology CRF.

**Author contributions**
J.S., Y.Z. and T.C. supervised the project. X.Y., X.M. and L.W. grew the $Fe_3GaTe_2$ and $Fe_3GeTe_2$ single crystals. S.G. conducted the SCXRD measurements and analyzed the structure. S.L. performed the STEM and LTEM characterizations. J.S., X.Y., and L.L. measured the magnetic properties of bulk crystals. C.B. and Z.L. prepared the thin-layer devices, and J.S. and X.Y. performed the transport measurements. J.S. and X.Z. conducted the RMCD and SHG measurements on the fresh $Fe_3GaTe_2$ nanoflakes. J.S., X.Y., Y.Z., and T.C. analyzed the results and wrote the paper with input from all authors.

**Competing interests**
The authors declare no competing interests.

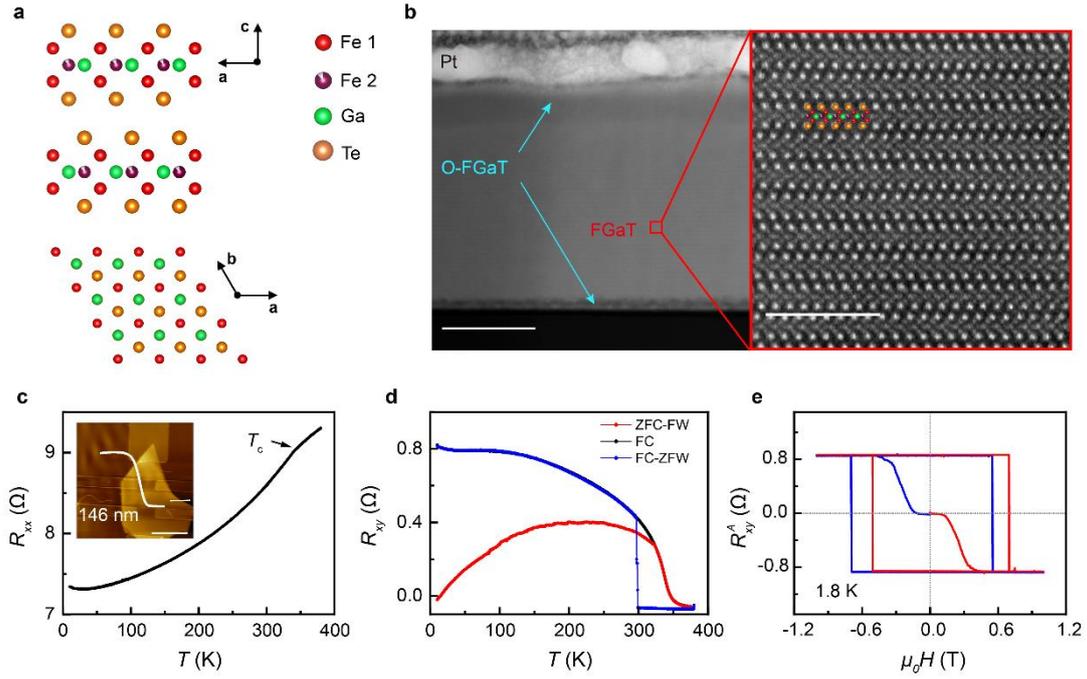

**Fig.1| Crystal structure and magneto-transport properties of thin Fe$_3$GaTe$_2$ flakes. a** Side and top views of the atomic lattice of a bilayer FGaT. **b** Cross-sectional STEM image of the FGaT nanoflake with a Pt capping layer. Scale bar is 70 nm. O-FGaT represents the surface oxidation. The atomic-resolution HAADF image of the red-boxed area in left panel is enlarged for clarity. Scale bar is 3 nm. **c** $R_{xx}$-$T$ curve of FGaT device #1 (146 nm). The inset shows its AFM topography and thickness. **d** $R_{xy}$-$T$ curves measured under different procedures. The applied magnetic field of FW (field-warming) and FC is 0.1 T. Here, $R_{xy}$ is symmetrized by ($R_{xy}$(+$H$)-$R_{xy}$(-$H$))/2 to remove the crossover of $R_{xx}$. **e** The ZFC AHE hysteresis loops measured at 1.8 K with different virgin magnetization directions.

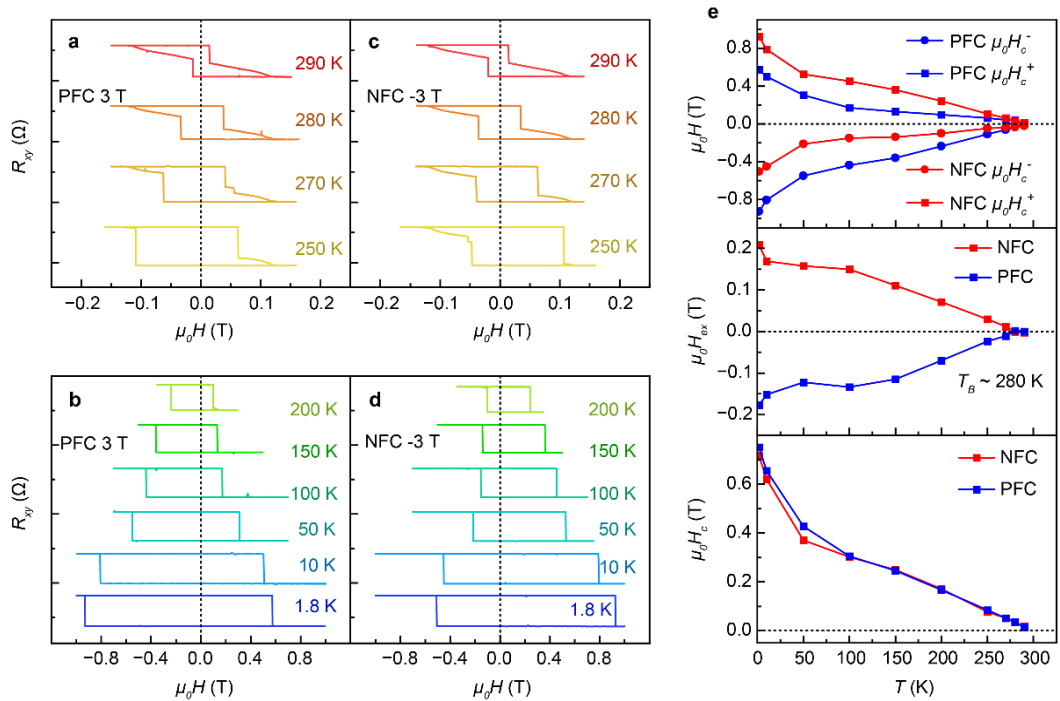

**Fig.2| Near-room-temperature EB phenomenon in Fe$_3$GaTe$_2$ thin-layer device. a-d** The AHE hysteresis loops of device #1 measured at various temperatures under positive (+3 T) and negative (-3 T) field cooling processes, respectively. **e** Temperature dependence of coercive field ($H_c^+$, $H_c^-$ and $H_c$) and bias field ($H_{ex}$), are summarized and plotted for both PFC and NFC procedures. The blocking temperature $T_B$ at which EB phenomenon disappears is approximately 280 K.

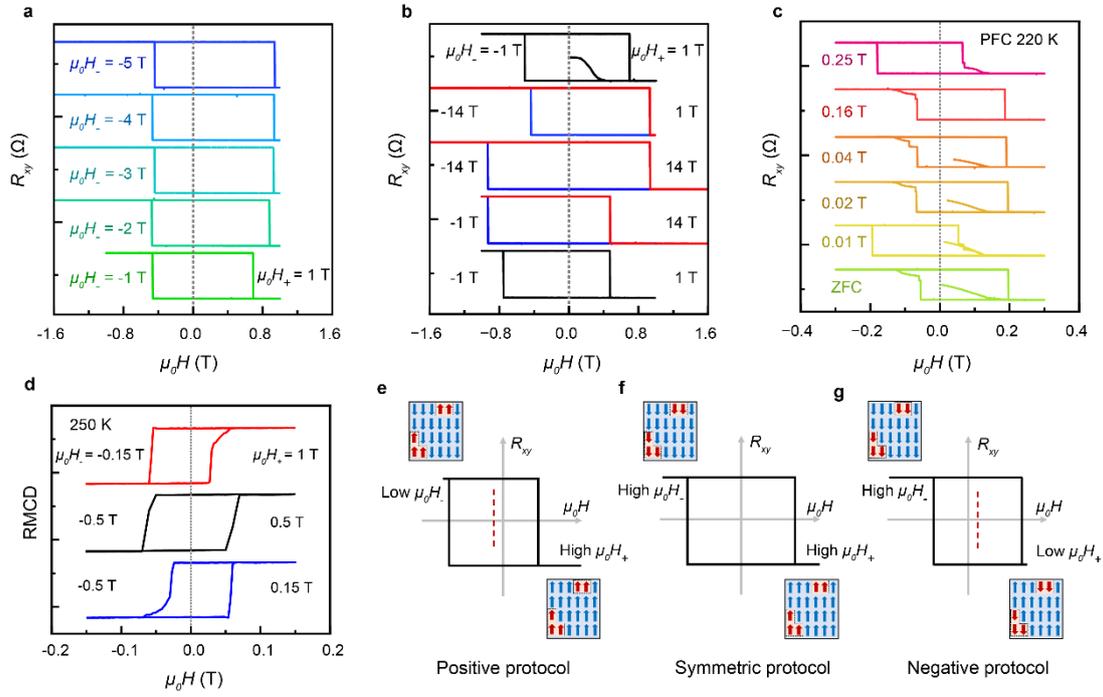

**Fig.3| The field-controllable EB and the proposed exchange spring model. a, b** The zero-field-cooling (ZFC) EB phenomenon are controlled through different field sweep ranges at 1.8 K. **c** The evolution of AHE hysteresis loops with the magnitude of cooling field at 220 K. **d** The RMCD hysteresis loops of the freshly exfoliated FGaT nanoflakes. **e-g** Schematic diagrams of the proposed exchange spring model measured by positive, symmetric, and negative protocols, respectively. The blue and red arrows represent the spin orientations of "soft" and "hard" ferromagnets, respectively.

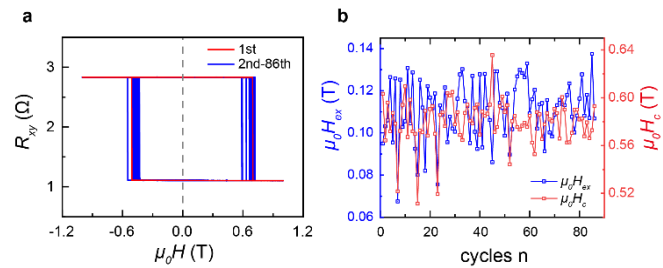

**Fig.4| The remarkable durability of EB. a** The training measurements of EB phenomenon with 86 repeated cycles at 1.8 K. **b** The statistical evolution of $H_{ex}$ and $H_c$ with repeated cycle times are summarized.

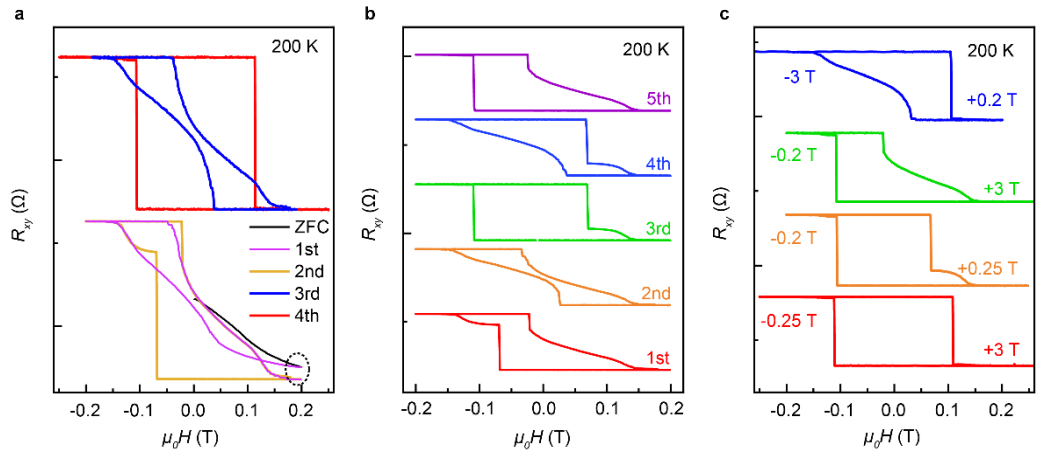

**Fig.5| The cumulative growth of minor loops and field-controllable EB at intermediate temperature. a** AHE hysteresis loops are consecutively measured by symmetric protocol with different field sweep ranges; 0.2 T for the first and second cycles; 0.19 T for the third cycle; 0.25 T for the fourth cycle. **b** Five representative hysteresis loops are selected from the 20 cycles measured by symmetric protocol of 0.2 T. **c** The symmetric and different asymmetric hysteresis loops can be selectively obtained by the specific field sweep ranges.